\documentclass[10pt]{article}
\usepackage{amsmath}
\usepackage[a4paper, margin=2cm]{geometry}
\usepackage[english]{babel}
\usepackage{graphicx}
\usepackage{float}
\usepackage{helvet}  

\usepackage{etoolbox}
\usepackage{wasysym}
\usepackage{gensymb}
\usepackage{url}
\usepackage{siunitx}
\usepackage{titlesec}
\titleformat{\paragraph}
{\normalfont\normalsize\bfseries}{\theparagraph}{1em}{}
\titlespacing*{\paragraph}
{0pt}{3.25ex plus 1ex minus .2ex}{1.5ex plus .2ex}

\patchcmd{\thebibliography}{\section*{\refname}}{}{}{}

\DeclareMathOperator*{\argmax}{arg max}

\date{\vspace{-10ex}}

\begin{document}

\title{Offset geometry for extended field-of-view in multi-contrast and multi-scale X-ray microtomography of lung cancer lobectomy specimens \\
}
\maketitle

\begin{center}
    \textbf{{\large }}
    \\
    
    {\large\textbf{Harry Allan}$^{1,2}$},
    \large{Adam Doherty}$^{1,2}$,
    \large{Carlos Navarrete-León}$^{1,2}$, 
    \large{Oriol Roche i Morgó}$^{1,2,}$\footnote[7]{Now with the Diamond Light Source, Harwell Science and Innovation Campus, Didcot, UK.}, 
    \large{Yunpeng Jia}$^{1,2,}$\footnote[8]{Now with the Department of Life Sciences, Birmingham City University, Birmingham, Uk.}, 
    \large{Charlotte Percival}$^{3}$,
    \large{Zoe Hagel}$^{3}$,
    \large{Kate E J Otter}$^{3}$,
    \large{Chuen Ryan Khaw}$^{3}$,
    \large{Kate Gowers}$^{3}$,
    \large{Helen Hall}$^{3,}$\footnote[9]{Now with the Department of Respiratory Medicine, King’s College Hospital, London, UK.},
    \large{Sam M Janes}$^{3}$,
    \large{Fleur Monk}$^{4}$, 
    \large{David Moore}$^{4,5}$, \large{Joseph Jacob}$^{3,6}$, \large{Marco Endrizzi}$^{1,2}$
    \\
    \textit{$^1$Department of Medical Physics and Biomedical Engineering, University College London, London, UK} \\

    \textit{$^2$X-ray Microscopy and Tomography Laboratory, The Francis Crick Institute, London, UK} \\

    \textit{$^3$UCL Respiratory, University College London, London, UK}

    \textit{$^4$Department of Cellular Pathology, University College London Hospitals NHS Foundation Trust, London, UK}

    \textit{$^5$CRUK Lung Cancer Centre of Excellence, UCL Cancer Institute, University College London, London, UK}
    
    \textit{$^6$Centre for Medical Image Computing, University College London, London, UK} \\

\end{center}

\begin{abstract}

X-ray microtomography is a powerful non-destructive technique allowing 3D virtual histology of resected human tissue. The achievable imaging field-of-view, is however limited by the fixed number of detector elements, enforcing the requirement to sacrifice spatial resolution in order to image larger samples. In applications such as soft-tissue imaging, phase-contrast methods are often employed to enhance image contrast. Some of these methods, especially those suited to laboratory sources, rely on optical elements, the dimensions of which can impose a further limitation on the field-of-view. We describe an efficient method to double the maximum field-of-view of a cone-beam X-ray microtomography system, without sacrificing on spatial resolution, and including multi-contrast capabilities. We demonstrate an experimental realisation of the method,  achieving exemplary reconstructions of a resected human lung sample, with a cubic voxel of 10.5 {\textmu}m linear dimensions, across a horizontal field-of-view of 4.3 cm. The same concepts are applied to free-space propagation imaging of a 2.7 mm segment of the same sample, achieving a cubic voxel of 450 nm linear dimensions. We show that the methodology can be applied at a range of different length-scales and geometries, and that it is directly compatible with complementary implementations of X-ray phase-contrast imaging.

\end{abstract}

\section{Introduction}

X-ray computed tomography (XCT), has proven to be an indispensable tool for the non-destructive 3D inspection of samples in a range of fields including medical sciences, materials engineering, and metrology \cite{withers2021x}. In particular, its higher-resolution implementation, X-ray microtomography (micro-CT) \cite{landis2010x} allows the probing of structures on the micro-scale. Access to 3D morphology has allowed detailed virtual histology of diseased human lung structure in cases such as fibrosis \cite{katsamenis2019x}, Covid-19 \cite{eckermann20203d}, and adenocarcinomas \cite{yi2021synchrotron}. Volumetric information from micro-CT enables 3D quantification of disease biomarkers \cite{kampschulte2013quantitative}, which can be demonstrably more sensitive than their 2D counterparts \cite{parameswaran2009three}, enhancing the understanding of disease and the effects of treatment.
\par
Conventional XCT reconstructs 3D distributions from many 2D projections formed predominantly by the attenuation of X-rays by the object in the beam. Yet despite its widespread adoption, the application of XCT to low-density materials, such as lung-tissue, still suffers from their inherently low X-ray attenuation cross section. X-ray phase-contrast imaging (XPCI) \cite{endrizzi2018x} is capable of achieving improvements in contrast-to-noise ratio (CNR) compared to attenuation-based imaging, and has thus been proposed as a solution to this problem. XPCI forms images based on phase-shifts imparted on the X-ray beam by the object, typically requiring beams with high levels of spatial and temporal coherence, such as those found at synchrotron radiation facilities \cite{snigirev1995possibilities,cloetens1999hard}. 
\par
Several techniques have been developed to enable the use of XPCI techniques with conventional sources, increasing their availability in laboratory and clinical settings. XPCI based only on free-space propagation (FSP) of the perturbed wavefront is achievable using laboratory sources \cite{wilkins1996phase}, however this comes at the cost of stringent requirements on source and detector characteristics. A number of other methods employ optical elements to alleviate some of these difficulties, including grating interferometry \cite{pfeiffer2006phase}, edge-illumination \cite{olivo2007coded}, beam-tracking \cite{vittoria2015beam}, and speckle-tracking \cite{zanette2014,quenot2021}. Each of these methods is able to recover independent attenuation and differential-phase signals, caused respectively by the imaginary ($\beta$) and real ($\delta$) components of the sample's complex refractive index. Alongside these quantities, they are also sensitive to the unresolved refraction and ultra-small angle X-ray scattering (USAXS) that leads to the dark-field contrast channel. Laboratory-based XPCI micro-CT has been utilised in a range of applications via a number of methods. These include, amongst many others,  small animal lung imaging \cite{weber2012investigation,velroyen2015grating,murrie2020real}, human breast imaging \cite{massimi2021detection,massimi2021volumetric,willner2014quantitative}, other animal model imaging \cite{savvidis2022monitoring,topperwien2017three,tapfer2012experimental}, and materials engineering \cite{shoukroun2020enhanced,gusenbauer2019porosity}.
\par
Techniques employing optical elements rely on either phase or attenuation effects to structure the beam in a way that creates phase sensitivity. The size of this optical element, alongside the size of the X-ray imaging detector, is typically fixed for any given system, therefore defining the usable imaging field-of-view (FOV). To overcome these limitations, it has become relatively routine at synchrotron sources to carry out offset scans, in which either the detector or rotation-axis is displaced and a 360\textdegree{} rotation is used to scan the entire sample \cite{kawata2010microstructural,walsh2021imaging}. In the case of a parallel synchrotron beam, opposing projections can be stitched together to create full-field equivalents, a method not consistent with the divergent geometry observed with most conventional sources. Alternatively, a simple linear weighting can be applied to overlap areas and reconstruction carried out across 360\textdegree{} \cite{vo2021data}. Offset scans have been demonstrated using grating interferometry with a synchrotron source \cite{mcdonald2009advanced}, where extended volumes were successfully reconstructed from truncated differential phase-contrast projections. 
\par
The same principles have also been demonstrated using cone-beam X-ray sources with an offset detector \cite{wang2002x,hansis2010iterative,sanctorum2021extended}. By maintaining alignment between the X-ray source, centre-of-rotation (COR), and the perpendicular yet-offset detector, a simple cone-beam geometry is maintained. This allows a straightforward processing by redundancy weighting of projections \cite{wang2002x}, followed by reconstruction using the Feldkamp Davis-Kress (FDK) algorithm \cite{feldkamp1984practical} or iterative methods \cite{hansis2010iterative, sanctorum2021extended}. Despite its wide adoption, such a geometry suffers a decrease in flux efficiency. Up to 50\% of the beam is lost as a direct result of offsetting the detector. Furthermore, for sources with fixed divergence, the minimum system length must be extended, to ensure that X-rays are incident upon the full extent of the offset detector.
\par
Alternatively, it is possible to laterally displace the COR with respect to the stationary source and detector. This way, there is no requirement to extend the system length and the full X-ray cone can be still captured by the detector. This allows an increase in the field of view, at no loss of X-ray flux density per detector element. To overcome the loss of the conventional cone-beam geometry, the acquired data may be rebinned onto a virtual detector, to approximate the offset-detector case \cite{lin2019reconstruction}. The rebinning process however increases computational complexity and may be a source of artefacts and inaccuracies. Instead, specific redundancy weights for the offset-COR case have also been derived and demonstrated using a flexible clinical gantry system \cite{belotti2023extension,belotti2022extension}. While this method was demonstrated using the combination of two complementary short (180\textdegree{} + cone angle) scans, the same information can also be captured using a single full 360\textdegree{} scan.
\par
We propose multi-contrast and multi-scale X-ray micro-CT with an offset COR geometry and a full 360\textdegree{} scan, to increase the achievable FOV in imaging of lung lobectomy specimens. We demonstrate the technique at multiple length scales, and show its compatibility with multiple contrast-mechanisms. A first demonstration utilises the beam-tracking technique to make the system sensitive to the effects of attenuation, phase, and dark-field, allowing quantitative separation of each contrast channel. Following this, the same technique is applied to a high-resolution setup at the same instrument, utilising FSP-based phase-contrast. Reconstruction is carried out using offset-COR redundancy weighting, followed by reconstruction without rebinning using a vector backprojection algorithm. The offset geometry allows up to 2x the FOV to be imaged without sacrificing pixel size, spatial resolution, or X-ray flux density per detector element. We show that this can be easily adapted to a wide range of existing XPCI and conventional systems.

\section{Materials and Methods}

\subsection{Beam-tracking X-ray phase-contrast}

\begin{figure}[]
    \centering
    \includegraphics[scale=0.87]{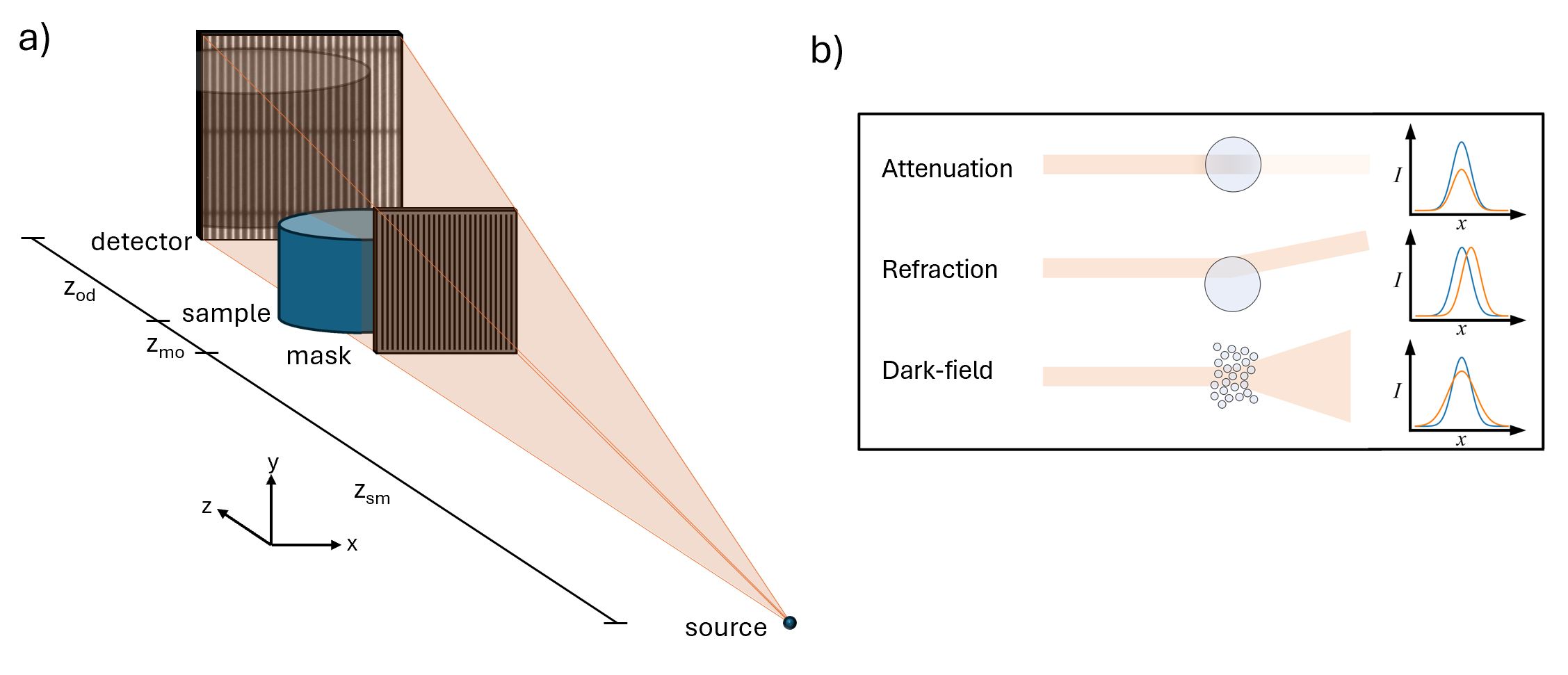}
    \caption{a) A diagram of the experimental system for beam-tracking x-ray microtomography with an offset geometry. The divergent x-ray beam is emitted from the source and is incident on an absorbing mask which structures the beam into an array of 1D beamlets. The beamlets interact with the sample and after some propagation distance z$_\text{od}$ are incident on a high-resolution detector capable of measuring the perturbations. The diagram is not to scale. b) The effects of different sample interactions for a single beamlet, with the corresponding 1D detector line profile. Each type of interaction may occur simultaneously, allowing the retrieval of three different contrast channels from a single image.}
    \label{BTgeometry_fig}
\end{figure}

Beam-tracking is a differential X-ray phase-contrast imaging technique that allows the
simultaneous extraction of attenuation, refraction, and dark-field signals in a single
shot \cite{vittoria2015beam}. An absorbing mask consisting of periodic apertures is used to structure the beam into a series of 1D (or 2D \cite{dreier2020tracking,navarrete2023x,lioliou2023laboratory}) beamlets, as illustrated in figure \ref{BTgeometry_fig}a. The principle of the beam-tracking technique is that sample-induced perturbations may be tracked by directly resolving the shape of the beamlets using a high-resolution detector. By comparing images acquired with and without the sample in the beam, these small perturbations can be quantitatively retrieved and converted into multi-contrast images. As the mask is placed on the source side of the sample, only photons contributing to image formation are incident on the sample, optimising the dose efficiency. While conceptually similar to several other techniques, including single-grid imaging \cite{morgan2011quantitative,wen2010single} and techniques based on tracking near-field speckle \cite{morgan2012x,berujon2012two}, beam-tracking utilises purpose built masks with small open fractions.

\par
The image retrieval process involves the extraction of horizontal 1D line profiles from the recorded projections, each corresponding to individual beamlets at indexed image positions $(x,y)$. As is illustrated in the inset of figure \ref{BTgeometry_fig}b, attenuation causes a decrease in the intensity of the beamlet, refraction causes an angular deflection of the beamlet, and dark-field causes a broadening of the beamlet. These are reflected as changes in the measured areas $A(x,y)$, centroids $\mu(x,y)$, and widths $\sigma(x,y)$ respectively of the extracted beamlet profiles upon the detector. Beamlet areas and widths are calculated numerically as the zeroth and second central moments of the measured beamlet intensity distributions. The shift in the centroid $\Delta\mu$ is estimated as the location of the maximum of the cross-correlation of the min-max normalised sample $I_S(x_i)$ and flat $I_F(x_i)$ beamlet intensity distributions

\begin{equation}
    \Delta\mu = \argmax_\epsilon
    \left( 
    \sum_i I_S(x_i)I_F(x_i+\epsilon) 
    \right),  
\end{equation}

where $\epsilon$ is the shift parameter, and $_i$ is the index along the 1D profile. The cross correlation is estimated first for integer $-w < \epsilon < w$, where $w$ is the beamlet half-width. This is followed by sub-pixel localisation by analytical parabolic interpolation of the maximisation landscape \cite{kashyap2025comparative}.
\par
Comparison of the flat and sample images allows extraction of the sample transmission, refraction angle, and dark-field signals by 

\begin{align}
    T(x,y) &= \frac{A_S(x,y)}{A_F(x,y)},
    \\
    \theta_R(x,y) &= \frac{\Delta\mu(x,y)}{\text{z}_\text{od}},
    \\
    \sigma^2(x,y) &= \frac{\sigma^2_S(x,y) - \sigma^2_F(x,y)}{\text{z}_\text{od}^2},
\end{align}

where z$_\text{od}$ is the propagation distance between the sample and detector, and the subscripts $_S$ and $_F$ refer to quantities calculated using images with and without the sample respectively.
\par
The retrieved transmission $T(x,y)$ is related to the line integral through the object $\int_{o}$ of the 3D distribution of the linear attenuation coefficient $\mu_0$ by 

\begin{equation}
    -\text{log}(T(x,y)) =  \int_{o} \mu_0 (x,y,z) \, dz .
    \label{logt_equation}
\end{equation}

The retrieved refraction angle is related to the derivative of the projected phase shift by $\theta_R(x,y) = \frac{1}{k} \, \frac{\partial \Phi(x,y)}{\partial x}$, where $\Phi(x,y)$ is the phase shift and $k$ is the wavenumber. Retrieving the projected phase through unidirectional integration allows the construction of the line integral

\begin{equation}
    -\frac{\Phi(x,y)}{k} = \int_{o} \delta (x,y,z) \, dz.
  \label{intref_equation}
\end{equation}

It has been shown previously that beam-tracking and edge-illumination retrieve a quantitative dark-field signal that is linear with thickness \cite{endrizzi2017x,doherty2023edge}, and thus we may finally also construct the line integral 

\begin{equation}
    \sigma^2 = \int_{o} \sigma^2_\phi (x,y,z)  \,   dz,
\end{equation}

where $\sigma^2_\phi$ is the linear scattering coefficient. It follows that each of the retrieved quantities is consistent with tomographic reconstruction, allowing the recovery of their 3D volumetric distributions.
\par
When fast scans are required, multi-contrast volumes may be reconstructed from a single image per projection angle, delivering spatial resolution limited by the mask period, yet still remaining quantitative \cite{hagen2014effect}. When higher spatial resolution is demanded, dithering may be employed, in which multiple images are acquired while the mask is translated in sub-period steps. The individual retrieved images are then stitched together in an interlaced pattern, leading to approximately aperture-driven resolution \cite{diemoz2014spatial}.

\subsection{Free-space propagation X-ray phase-contrast}

X-ray phase-contrast imaging based on free-space propagation is an established technique in which simple propagation of a sufficiently coherent X-ray beam converts phase-shifts into intensity modulations \cite{paganin2006coherent}. The intensity incident upon a detector placed $z_\text{od}$ behind the sample can be approximated by the transport of intensity equation (TIE) as \cite{wilkins1996phase}

\begin{equation}
    I_d(x,y) = \frac{I_s(x,y)}{M^2} \left(
    1 + \frac{z_\text{od}}{k M} \nabla_\bot^2 \phi (x,y;k)
    \right),
    \label{fsp_equa}
\end{equation}

where $I_s(x,y)$ is the intensity immediately after leaving the sample, $M$ is the geometric magnification of the system, and $\nabla_\bot^2 \phi (x,y;k)$ is the transverse Laplacian of the object-induced phase shift projected onto the $xy$-plane. Thus, assuming sufficient resolving power, for $z_\text{od} > 0$ the resultant image contains both phase and attenuation induced contrast. Under the assumption of a homogenous object, an approximate phase distribution compatible with tomography can be retrieved from the mixed images using an inversion of the TIE \cite{paganin2002simultaneous}. While being a widely popular technique at synchrotron radiation facilities, FSP based imaging can also be applied to laboratory-based systems, employing either high-resolution detectors \cite{esposito2025laboratory} or small X-ray focal spots \cite{eckermann2020phase}.

\subsection{Offset Geometry and Redundancy Weighting}

\begin{figure}[]
    \centering
    \includegraphics[scale=0.85]{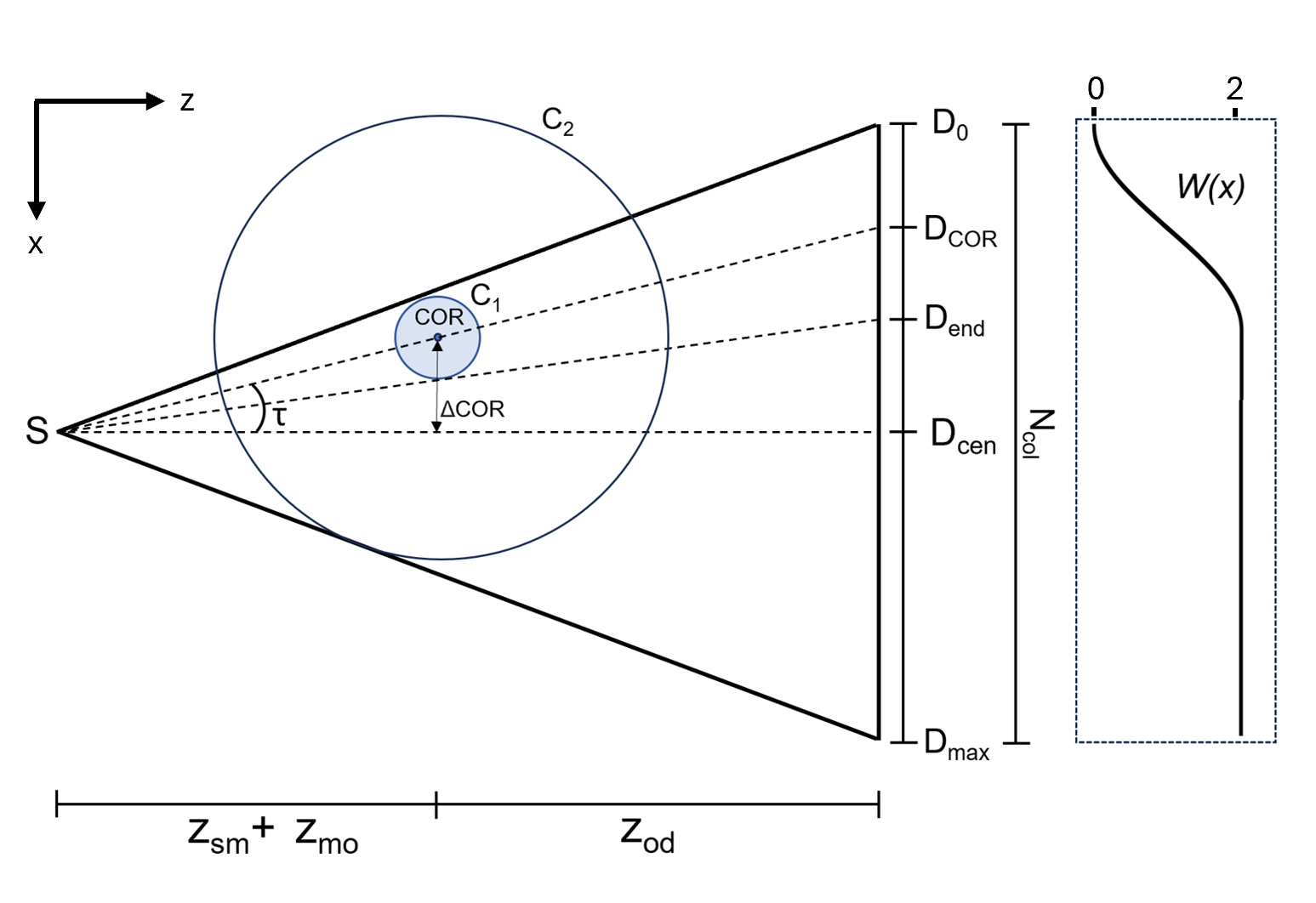}
    \caption{A diagram demonstrating the definitions of the offset tomography geometry. The centre-of-rotation (COR) is offset by an amount $\Delta$COR from the axis between the source and perpendicular detector, creating an angle $\tau$ with the detector centre. Only the region of the sample enclosed by the circle C$_1$ remains with the field-of-view throughout a 360\textdegree{} rotation of the sample. The remaining region enclosed by C$_2$ is seen for some period $<$360\textdegree{} and thus must be weighted differently during the reconstruction. W(x) illustrates the weighting function for the demonstrated geometry.}
    \label{offsetgeometry_fig}
\end{figure}

Consider a cone-beam beam geometry with its origin at the centre-of-rotation COR, as illustrated in figure \ref{offsetgeometry_fig}. In the sample reference frame, the circular tomographic trajectory with projection angles $\alpha$ is described by the relative positions of the source and (centre of) detector, indicated by $\mathbf{S}(x,y,z)$ and $\mathbf{D_{cen}}(x,y,z)$, respectively. For brevity, the coordinate dependence $(x,y,z)$ is henceforth omitted. To define these positions, by the conventions of the Astra Toolbox \cite{van2016fast}, we use the opposing unit vectors $\mathbf{V_s} = [\text{sin}(\alpha), \, 0, \, -\text{cos}(\alpha)]$, and $\mathbf{V_d} = [-\text{sin}(\alpha), \, 0, \, \text{cos}(\alpha)]$, where $\alpha$ is the angle of a given projection. We thus can write the source and detector positions as

\begin{equation}
    \mathbf{S} = (\text{z}_\text{sm} + \text{z}_\text{mo})\mathbf{V_s}  + (\Delta\text{COR})\mathbf{U},
\end{equation}

\begin{equation}
    \mathbf{D_{cen}} = (\text{z}_\text{od})\mathbf{V_d}  + (\Delta\text{COR})\mathbf{U},
\end{equation}

where $\Delta$COR is the offset of the COR along x, $\text{z}_\text{sm}$ is the source-to-mask distance, $\text{z}_\text{mo}$ is the mask-to-object distance, and $\mathbf{U} = [\text{cos}(\alpha), \, 0, \, \text{sin}(\alpha)]  $ is the unit vector describing the detector pixel orientation, perpendicular to the line formed by $\mathbf{SD_{cen}}$. From figure \ref{offsetgeometry_fig} it can be seen that only the region enclosed by the circle C$_1$ remains within the FOV for the duration of a 360\textdegree{} rotation of the sample about COR. Meanwhile, regions enclosed by the circle C$_2$ but not by C$_1$ remain within frame for some subset of the  $<$360\textdegree. A redundancy weighting must be applied along $x$ to the acquired projections, to correct for the increased density of rays counted within C$_1$ compared to C$_2$. For offset detector geometries, a redundancy weighting function $W(x)$ that is spatially symmetric about the COR must fulfill $W(\text{D}_\text{COR} + n$) + $W(\text{D}_\text{COR} - n$) = 2 \cite{cho1996cone}, where the points $\text{D}_\text{COR} + n$ and $\text{D}_\text{COR} - n$ are detector positions symmetric about the projected COR $\text{D}_\text{COR}$. However, for the case of an offset COR geometry, with COR offset angle $\tau = \tan^{-1}(\Delta\text{COR} / (\text{z}_\text{sm} + \text{z}_\text{mo}) )$, this must be adjusted to a symmetry in angle $\beta$ about $\text{D}_\text{COR}$, $W(\text{D}_\text{COR} + \beta$) + $W(\text{D}_\text{COR} - \beta$) = 2. The function should furthermore be absent of discontinuities and offer a smooth transition between the maximum and minimum values. We fulfill these criteria by adopting a similar smooth weighting function as previously used for complementary offset-COR short scans \cite{belotti2023extension}, but instead we apply this principle to a full 360\textdegree{} rotation. We describe this as a function of the detector position $x$ as

\begin{equation}
W(x) = 
\begin{cases}
    0 & , \, \text{if } x < D_0 \\

    \text{sng}\{\tau\}
    \sin 
    \left(
    \frac{\pi}{2} \frac{\tan^{-1}(x/R) + \tau }{\tan^{-1}(N_\text{cols}/(2R)) - |\tau|  }
    \right) 
    
    & , \, \text{if } D_0 < x \leq D_\text{end} \\

    2 & , \, \text{if } x > D_\text{end}
\end{cases},
\end{equation}

where $N_\text{cols}$ is the number of detector columns, and $R = \text{z}_\text{sm} + \text{z}_\text{mo} + \text{z}_\text{od} $. The size of the redundant region is given by $ |D_0 D_\text{end}| = | N_\text{cols}/2 + R \tan\{ \tan^{-1}(N_\text{cols}/(2R)) - 2| \tau  |  \} | $. An example weighting function is plotted inline with the corresponding detector positions in figure \ref{offsetgeometry_fig}.

To reduce truncation artefacts \cite{wang2002x,sanctorum2021extended} without computationally expensive iterative methods \cite{sanctorum2021extended}, we apply the redundancy weighting after first applying the ramp filter $\hat{f_R}(\omega)$ to the projections \cite{cho1996cone}. Similar to the sinogram extension method \cite{cho1996cone}, artefacts due to discontinuities are avoided by padding images with a copy of the width reflected sinogram on the truncated side, and edge values (normally zeros) on the non truncated side, prior to Fourier transformation. The padded values are then removed before reconstruction. The reconstruction procedure is described by

\begin{equation}
    V(x,y,z) = \text{BP} \left( W(x) \times \mathcal{F}^{-1}\{ 
    \hat{f_R}(\omega) \times \mathcal{F} [ P(x,y;\alpha)   ]
    \}
    \right),
\end{equation}

where BP denotes the backprojection operation that reconstructs the volume $V(x,y,z)$ from projections $P(x,y;\alpha)$, and $\mathcal{F}$ and $\mathcal{F}^{-1}$ are the forward and inverse Fourier transform operations respectively. The backprojection step is implemented using a vector backprojection algorithm \cite{van2016fast}, allowing the exact custom geometry to be defined, and (alongside the application of specific offset-COR weights) removing the requirement of a rebinning step. Easily adaptable code to calculate and apply offset-geometry weights, alongside a demonstration on simulated data, is made publicly available \cite{harry_code_offset}.

\subsection{Simulations}

The accuracy of the weighting and reconstruction method was first investigated via a numerical simulation study. The central detector row corresponding to an offset geometry such as that shown in figure \ref{offsetgeometry_fig} was defined using a series of vectors, and a truncated sinogram simulated by forward projection of the Shepp-Logan phantom using the projectors of the Astra Toolbox \cite{van2016fast}. The geometry parameters were chosen to model the demonstrated experimental system. Parameters were chosen as: $\text{z}_\text{sm} + \text{z}_\text{mo} =$ 
870 mm, $\text{z}_\text{od} = $ 170 mm, pixel size p$_x =$ 12.5 \textmu m (to match the demagnified dithering step), and $\Delta\text{COR} =$ 7.6 mm. A total of 2701 projections across 360\textdegree{} were simulated, upon a detector of 2424 pixel width. Simple X-ray line integrals were simulated without considering phase effects. For comparison, a large detector conventional scan was also simulated, in which the number of pixels were doubled to 4848, and the offset was reduced to zero.
\par
Simulated sinograms were then reconstructed using the described redundancy weighting and the vector cone-beam geometry in the Astra toolbox. The mean squared error (MSE) between the reconstructions and ground-truth were then calculated and used for quantitative comparison.

\subsection{Experiments}

The system was utilised to image a section of resected human lung tissue of approximately 4 x 4 x 2 cm$^3$. The sample was obtained with patient consent from a left upper lobectomy at University College London Hospital, in a patient diagnosed with squamous-cell carcinoma of the lung. The large sample offered an opportunity to demonstrate the possibility of imaging large biological objects while maintaining a high spatial resolution. The sample was initially fixed in 70\% ethanol, followed by dehydration in a series of ascending ethanol concentrations (70\%, 80\%, 90\%, 100\%) for at least 24 hours per stage. The sample was then placed in a bath of hexamethyldisilazane for 4 hours, to minimise structural changes that would occur during the final stage of overnight air drying under a fume hood \cite{mai2017thin,pallares2022tissue}. Some tissue shrinkage was observed during the ethanol dehydration stage. Following this, minimal changes occurred during the air drying. For imaging with beam-tracking, the sample was placed inside of a sealed polypropylene container. For the high-resolution FSP experiment, a 2.7 mm diameter segment of the same sample was removed, and mounted on top of a skewer. 
\par
Experiments were carried out using the NXCT (National Research Facility for lab-based X-ray Computed Tomography) multi-contrast X-ray micro-CT system \cite{i2024new}. The system features a Rigaku MicroMax 007-HF rotating molybdenum anode X-ray source, which was operated at 50 kV tube voltage and 24 mA current, with a focal spot size of 70 \textmu m. Two exit windows allow simultaneous imaging on two different end-stations. For the beam-tracking experiment, the beam was filtered with a total of 0.18 mm of aluminium, resulting in a mean energy of approximately 23 keV, as estimated using the SpekPy simulation package \cite{poludniowski2021spekpy}. A custom-built lens-coupled scintillator detector, with an effective pixel size of 15 \textmu m and a FOV of $\sim$ 30 mm, was integrated into the system. The X-ray beam was structured using a 1D absorption mask with apertures of 10 \textmu m, and a period of 79 \textmu m. The absorbing septa of the masks were created by depositing 120 \textmu m of gold onto a silicon substrate for a total mask thickness of 400 \textmu m. The system geometry was set to $\text{z}_\text{sm} = $ 820 mm, $ \text{z}_\text{mo} =$ 50 mm, and $\text{z}_\text{od} = $ 170 mm. This resulted in a native planar FOV at the sample plane of 2.5 cm. For tomography, the COR was offset by $\Delta$COR = 7.6 mm. This resulted in an increase of the horizontal FOV by a factor of 1.7x up to 4.3 cm, sufficient to cover the entire sample. 
\par
Projections were acquired at 2701 angles equally spaced over 360\textdegree, with a dithered acquisition to improve spatial resolution. To implement this efficiently, a separate tomography was acquired at each dithering step, with the mask displaced in steps of 9.9 \textmu m. An exposure time of 3 s per projection resulted in a total exposure time of 18 hours. Attenuation, refraction, and dark-field contrasts were retrieved from each projection, followed by interlaced stitching. The process of dithering results in unequal pixel lengths of retrieved images in the x and y directions, which was corrected using 3rd order spline interpolation to yield square pixels of 10.5 \textmu m x 10.5 \textmu m, matching the demagnified dithering step. This maintains the aperture driven resolution in the x-direction, with the y resolution still being limited by the detector and magnified source point spread functions. The retrieved and stitched projections were filtered by a Gaussian filter of standard deviation 0.7 pixel, as a compromise to increase CNR without over-smoothing image details. Phase integration was implemented using the cumulative sum of retrieved refraction angle images, in the direction of phase sensitivity. As the refraction images are only truncated on one side, the area outside of the sample provides air as a known boundary condition to constrain the phase integration.
\par
For the FSP-based experiment, the same source settings were utilised on the high-resolution end-station, without additional filtration. A different custom-built lens-coupled scintillator detector, with an effective pixel size of 450 nm and a horizontal FOV of 1.44 mm, was integrated into the system. The detector was placed 390 mm away from the source, with the propagation distance set to $\text{z}_\text{od} = $ 10 mm. For tomography, the COR was offset by $\Delta$COR = $\sim$ 630 \textmu m. This resulted in an increase of the horizontal FOV by a factor of 1.85x up to $\sim$ 2.7 mm.
\par
Projections were acquired at 5001 angles equally spaced over 360\textdegree. An exposure time of 5 s per projection resulted in a total exposure time of just under 7 hours. Projections were processed using Paganin single-distance phase retrieval \cite{paganin2002simultaneous} with $\frac{\delta}{\beta}$ = 1200.
\par
Sinograms were processed with a stripe removal algorithm to reduce the appearance of ring artefacts. Briefly, stripes were identified by averaging sinograms along the angle direction and subtracting the low-pass filtered component. The resulting profile was then used to normalise the sinograms by subtraction. Individual volumes were reconstructed using the described redundancy weighting procedure, and vector backprojection using the Astra toolbox. The beam-tracking volumes were post-processed with a 3D Gaussian filter of standard deviation 1 pixel, again chosen as a compromise to enhance CNR while not over-smoothing image details.

\section{Results}

\subsection{Simulations}

\begin{figure}[h]
    \centering \includegraphics[scale=0.85]{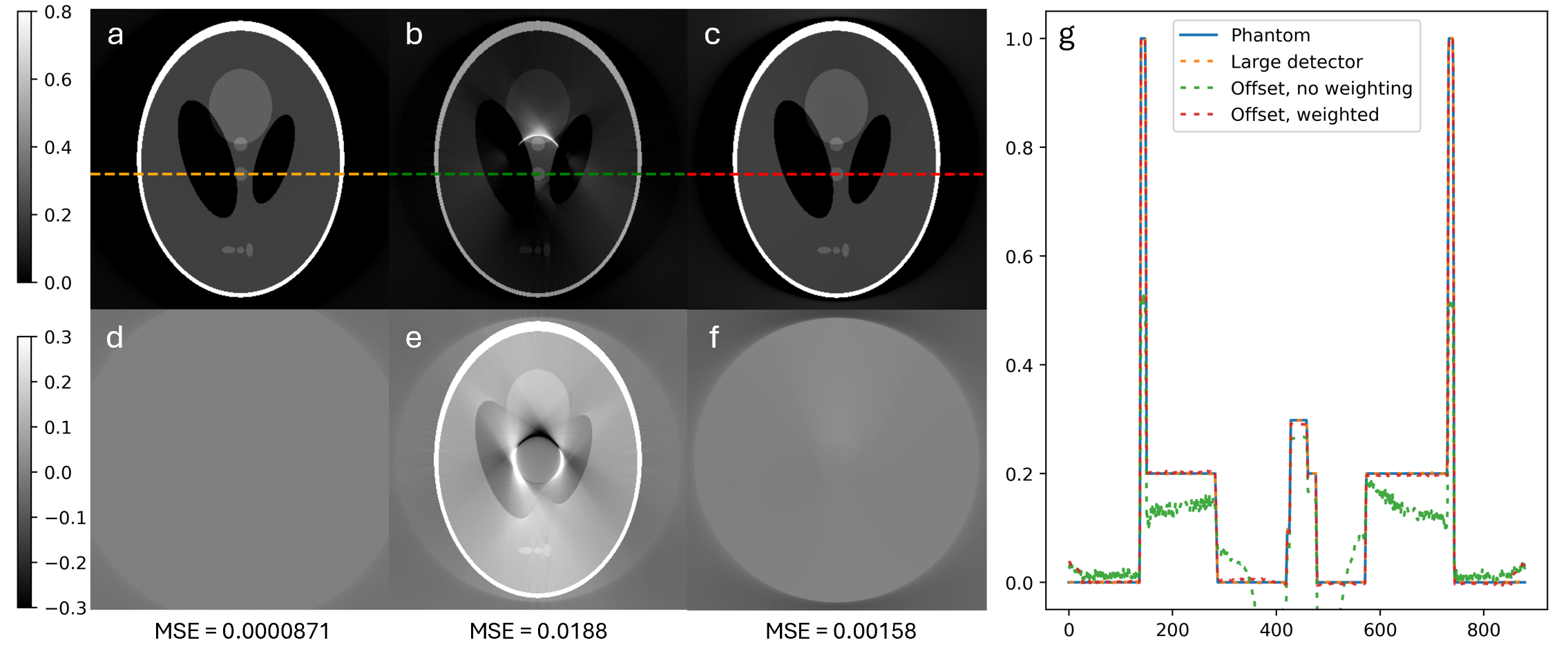}
    \caption{Results of the offset scan simulation study illustrating the reconstructions achieved using the large detector conventional scan (a), the offset COR without redundancy weighting (b), and the offset COR with redundancy weighting (c). The bottom row indicates the difference images between the reconstructions and the ground truth, annotated with the mean squared error (MSE). The annotated dashed lines (a,b,c) are plotted in comparison to the ground truth (g). Within the sample region, the redundancy weighted offset scan provides comparable results to the large detector conventional scan, and is consistent with the ground truth.}
    \label{sim_results_fig}
\end{figure}

The results of the simulation study are illustrated in figure \ref{sim_results_fig}. Reconstructions of the large detector conventional scan (\ref{sim_results_fig}a), offset COR with no weighting (\ref{sim_results_fig}b), and offset COR with redundancy weighting (\ref{sim_results_fig}c) are displayed on the top row. The corresponding difference images between the reconstructions and ground-truth are displayed on the bottom row. The dashed lines annotated on figures \ref{sim_results_fig}a, \ref{sim_results_fig}b, \ref{sim_results_fig}c are plotted on figure \ref{sim_results_fig}g. As expected, reconstruction of the large detector conventional scan accurately reproduces the phantom with a minimal MSE. While the offset COR with no weighting does reconstruct the approximate morphology of the sample, the unweighted redundant rays produce significant artefacts at the centre of the phantom. Furthermore, there is a large loss in quantitativeness, particularly visible in the profiles of figure \ref{sim_results_fig}g. The offset COR scan with redundancy weighting accurately reconstructs the phantom, with the line profile matching the ground truth, and with a much reduced MSE compared to the no weighting case. While the MSE is still higher than the large detector case, most error occurs outside of the valid reconstruction region. It is thus shown that an offset COR scan combined with redundancy weighting can produce comparable reconstructions to a scan acquired with a detector twice as large.

\subsection{Experiments}

\begin{figure}[]
    \centering
    \includegraphics[scale=0.51]{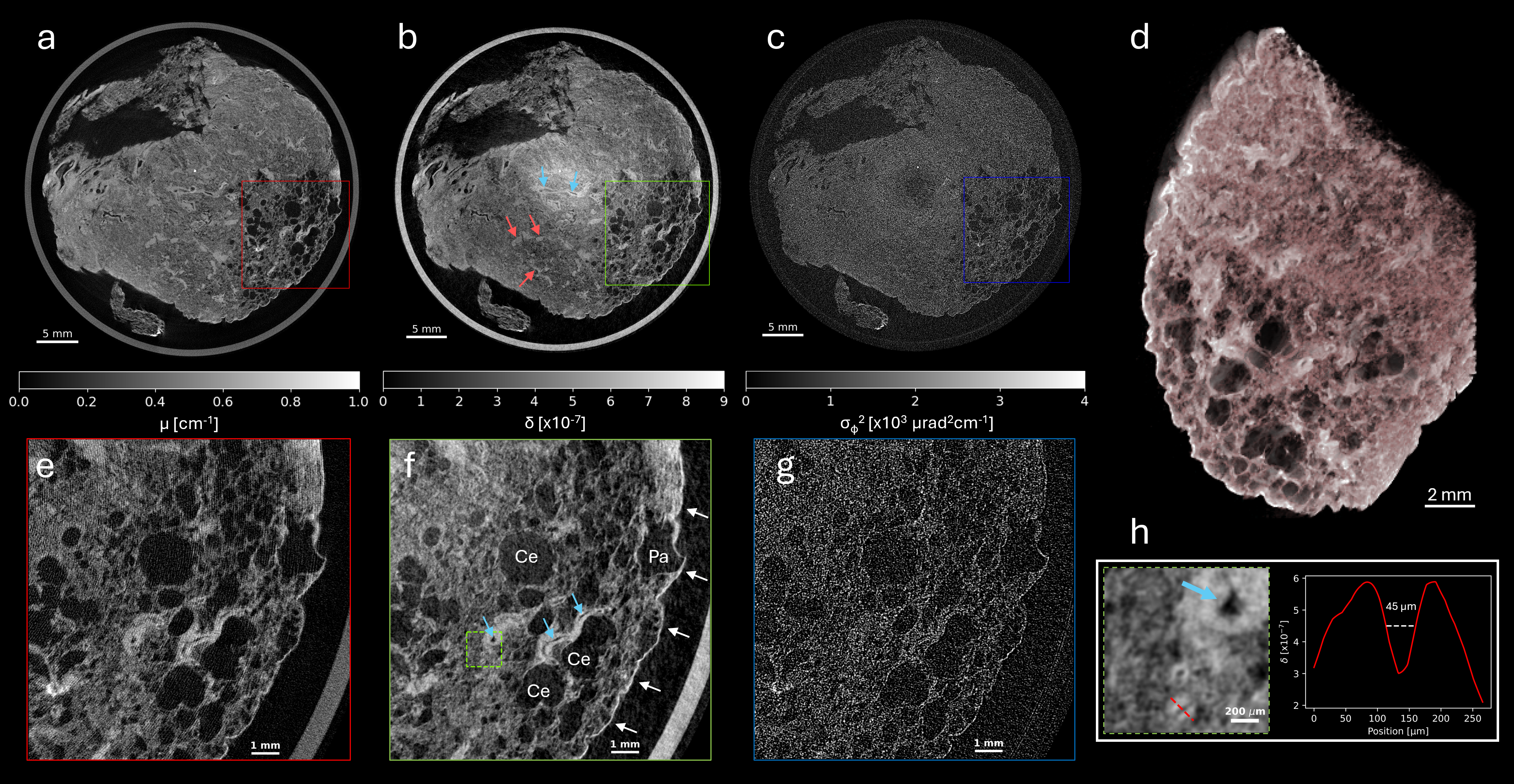}
    \caption{Axial slices of the offset beam-tracking tomography in attenuation (a), phase (b), and dark-field (c) contrast channels. Volumetric information allows the network of arteries (blue arrows) and veins (red arrows) to be traced through the tissue. Zoomed views at the positions indicated by the coloured boxes also display the attenuation (e), phase (f), and dark-field (g) channels, where it is evident that the phase channel enables the visualisation of fine, low-contrast features. The phase zoom (f) illustrates the well defined pleura (white arrows) and examples of fine pulmonary arteries (blue arrows), as well as markers of disease: enlarged  air spaces indicative of centrilobular (Ce) and paraseptal (Pa) emphysema. A further zoom (h), taken from the position indicated by the dashed green box (f), demonstrates the ability to resolve fine features. A line profile is taken through an arterial lumen of several tens of micrometres, which can be traced back to the larger vessel indicated by the red arrows in b. The region in f is rendered as a volume (d), where many large emphysematous air spaces are made visible.}
    \label{results_fig}
\end{figure}

Figure \ref{results_fig} illustrates the reconstruction of the beam-tracking scan using the proposed method. Axial slices of the attenuation (\ref{results_fig}a), phase (\ref{results_fig}b), and dark-field (\ref{results_fig}c) contrast channels are shown, alongside corresponding zooms (\ref{results_fig}e, \ref{results_fig}f, \ref{results_fig}g). Volumetric information allows the tracing of pulmonary vasculature, with both venous (red arrows) and arterial (blue arrows) systems labelled in figure \ref{results_fig}b. Fine, low-contrast details are made more easily visible in the phase channel, with a number of these details annotated on figure \ref{results_fig}f. In particular, a network of fine vasculature can be seen, which can be followed back through the volume to the vessels labelled in figure \ref{results_fig}b. A further zoom and line profile (\ref{results_fig}h) demonstrates an arterial lumen with a full-width-at-half-maximum (consisting of both the lumen width and the system spatial resolution) of 45 \textmu m. Note that for this sample, much of the venous system appears with a clear air-filled lumen, likely as a consequence of the sample preparation protocol. A number of enlarged air spaces indicative of centrilobular (Ce) and paraseptal (Pa) emphysema are visible in the slices, and particularly also in the volume rendering (\ref{results_fig}d). These lesions are both markers of disease that are consistent with the patient's smoking history.

\begin{figure}[]
    \centering
    \includegraphics[scale=0.75]{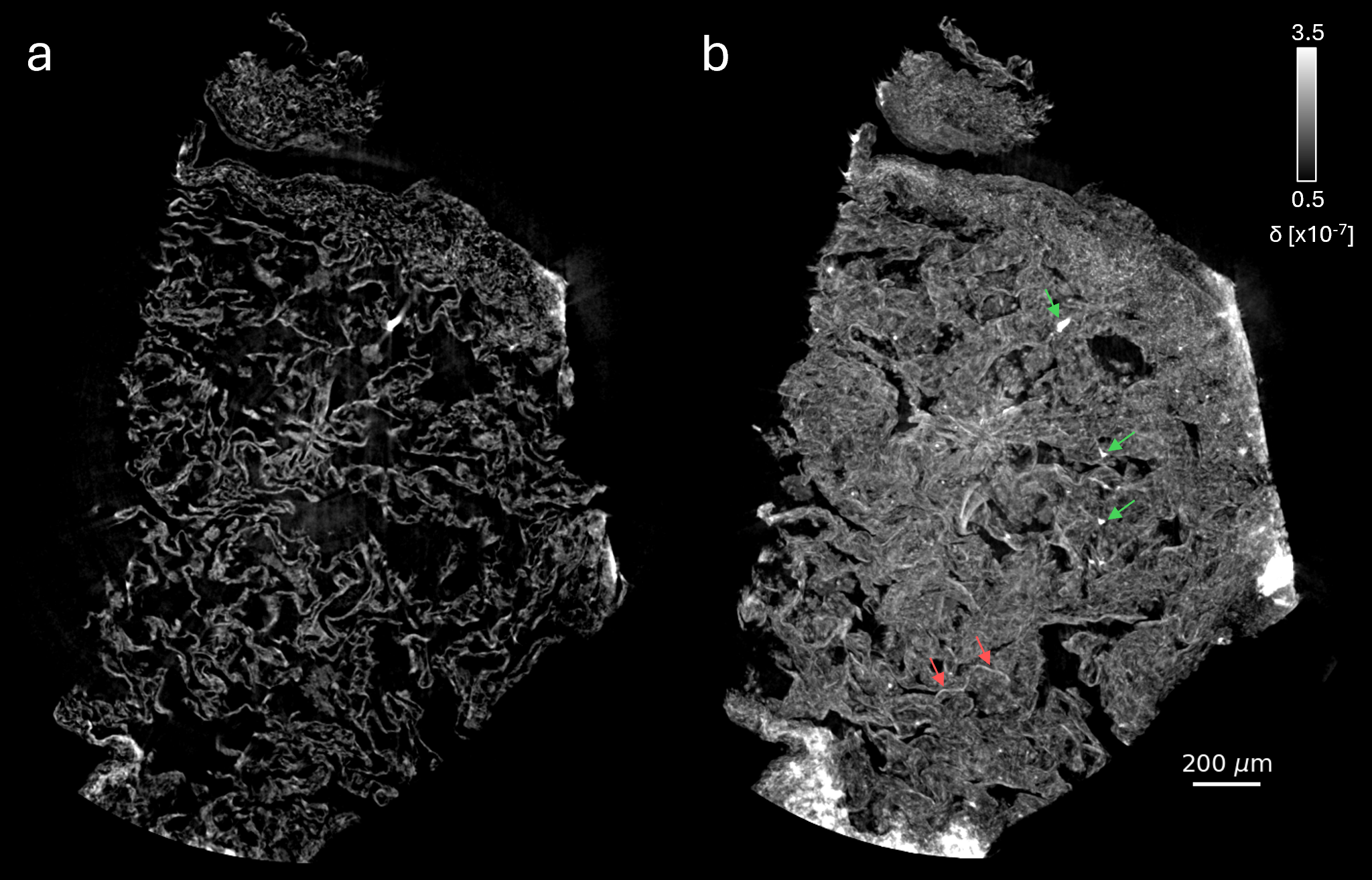}
    \caption{Axial slice (a) and maximum intensity projection corresponding to a thickness of 67.5 \textmu m (b) of the offset high-resolution scan. The axial slice demonstrates the complex mesh of alveolar septa that make up the lung's gaseous exchange system. The maximum intensity projection particularly highlights a number of calcifications (green arrows), and the network of vessels (red arrows) that run alongside the alveoli.}
    \label{muscope_results_fig}
\end{figure}

Figure \ref{muscope_results_fig} illustrates the reconstruction of the high-resolution scan using the proposed method. The axial slice in figure \ref{muscope_results_fig}a shows how at higher resolution, the collapsed and closely spaced alveolar septa visible in figure \ref{results_fig} are now better defined. The maximum intensity projection in figure \ref{muscope_results_fig}b highlights brighter structures including calcifications (green arrows) and a number of small vessels (red arrows) with a diameter of $\sim$ 8 \textmu m, which could represent capillaries, and which are smaller than the beam-tracking voxel size of 10.5 \textmu m. This demonstrates the varying morphological information that is attainable through multi-scale imaging.

\section{Discussion}

We have demonstrated through simulations and experiments a method to extend the native FOV of multi-contrast and multi-scale XPCI systems. An extension factor of up to 1.85x was shown experimentally, though this could be increased up to 2x if the sample requires it. This allowed the imaging of a resected tissue sample larger than the native FOV, without sacrificing spatial resolution. The large FOV of the beam-tracking system, combined with resolving power for vessels of tens of micrometres will enable advanced applications, such as the segmentation and tracing of airway and vascular networks \cite{pakzad2024evaluation}. The context provided by the large FOV scan can work synergistically with the high-resolution scan, where the same technique was applied. The possibility to investigate different length scales using the same instrument could allow selective zooming of areas chosen based upon the large FOV overview scan. 
\par
Of particular value is that the redundancy weighting procedure allows the reconstruction of accurate volumes, without the necessity to use iterative methods. These are associated with lengthy computation times and may become unusable in the case of the especially large experimental volumes shown. This factor is of particular importance if the technique is applied in fields where rapid reconstructions are required, such as interoperative imaging, for which XPCI has already shown positive outcomes \cite{massimi2021detection}. The ability to tailor FOV to the particular sample, without sacrificing spatial resolution, would be highly beneficial in such scenarios, where it may not be known in advance of the surgery the exact sample size that is to be expected.
\par
While the multi-contrast capabilities were demonstrated using a beam-tracking approach, the same principles can be extended to a range of multi-contrast XPCI systems such as grating-interferometry \cite{pfeiffer2006phase}, edge-illumination \cite{olivo2007coded}, and speckle-tracking \cite{zdora2018state}. In particular, it should be noted that offset scanning does not require any specialised adaptions other than the ability to translate the rotation stage, and thus is inherently compatible and stackable with FOV-extension methods such as tiled-gratings \cite{meiser2016increasing,schroter2017large} and scanning-based methods \cite{astolfo2017large,willer2018x}. This has the potential to increase the range of samples across which lab-based multi-contrast XPCI may be applied, and in particular may be a valuable tool in recent large-FOV developments such as human scale phase-sensitive imaging \cite{viermetz2022dark}. 
\par
The quasi-parallel beam high-resolution system featured a vastly different geometry and a different image formation mechanism to the beam-tracking system, further demonstrating that the technique is generalisable to a range of systems. This could open application to other methods in which a cone-beam reconstruction is required, including advanced multi-modal imaging \cite{brombal2023}, and even holotomography using focused synchrotron radiation \cite{eckermann20203d}. 

\section{Conclusion}

We have demonstrated a method by which the field of view of existing lab-based X-ray phase-contrast imaging micro-CT systems may be extended, without sacrificing spatial resolution. This makes it possible to overcome the limits of detector and optical element sizes in the push to image larger samples, or similarly may be used to image a given sample using a higher-resolution system. We have shown that this method is compatible with dark-field and phase-contrast channels, despite the differential nature of the phase signal requiring an additional integration step before reconstruction. We also showed that the approach is straightforward to adapt to other lab-based X-ray phase-contrast imaging systems, such as those based on free-space propagation, and that it can be directly applied at different length scales. These results are translatable to other phase contrast systems utilising optical elements, as well as to conventional micro tomography systems, thus offering an effective way to push beyond the limits imposed on sample size in conventional micro-CT. 

\section{Acknowledgements}

This work is supported by the EPSRC-funded UCL Centre for Doctoral Training in Intelligent,
Integrated Imaging in Healthcare (i4health) (EP/S021930/1), the Department of Health’s
NIHR funded Biomedical Research Centre at University College London Hospitals, and the
National Research Facility for Lab X-ray CT (NXCT) through EPSRC grants EP/T02593X/1
and EP/V035932/1; and by the Wellcome Trust 221367/Z/20/Z. This work is also supported by the Francis Crick Institute, which receives its core funding from Cancer Research UK (CC0102), the UK Medical Research Council (CC0102), and the Wellcome Trust (CC0102). SMJ is supported by CRUK programme grant (EDDCPGM/100002), and MRC Programme grant (MR/W025051/1). SMJ receives support from the CRUK Lung Cancer Centre of Excellence (C11496/ A30025) and the CRUK City of London Centre. This work was partly undertaken at UCLH/UCL who received a proportion of funding from the Department of Health’s NIHR Biomedical Research Centre’s funding scheme.
\par
We are grateful for the assistance of Alberto Astolfo for laboratory
management and support.
\par
Approval of all ethical and experimental procedures and protocols was granted by South Central Hampshire B Research Ethics Committee
(REC number: 18/SC/0514).

\end{document}